\documentclass[10pt,letterpaper,onecolumn]{article}

\usepackage{balance}
\usepackage{comment}
\usepackage{graphicx}
\usepackage[justification=centering,small]{caption}

\date{September 8, 2017}
\usepackage{algpseudocode}
\usepackage{algorithm}
\usepackage{mathtools}

\begin{document}
\title{TECHNICAL REPORT\\~\\Adaptive Rate Allocation for View-Aware Point-Cloud Streaming}
\author{Mohammad Hosseini\\University of Illinois at Urbana-Champaign (UIUC)}
\maketitle
\section{Large-Scale View-Aware Adaptation}
In the context of view-dependent point-cloud streaming in a scene, our rate allocation is ``adaptive'' in the sense that it priorities the point-cloud models depending on the camera view and the visibility of the objects and their distance as described. The algorithm delivers higher bitrate to the point-cloud models which are inside user's viewport, more likely for the user to look at, or are closer to the view camera or, while delivers lower quality level to the point-cloud models outside of a user's immediate viewport or farther away from the camera. For that purpose, we hereby explain the rate allocation problem within the context of multi- point-cloud streaming where multiple point-cloud models are aimed to be streamed to the target device, and propose a rate allocation heuristic algorithm to enable the adaptations within this context. To the best of our knowledge, this is the first work to mathematically model, and propose a rate allocation heuristic algorithm within the context of point-cloud streaming.

The rate selection and allocation problem is the well-known binary Knapsack optimization problem, for which one approach to tackle is to transmit a subset of the whole point clouds within the 360-degree environment. The binary Knapsack problem is NP-hard, but efficient approximation algorithms can be utilized (fully polynomial approximation schemes), so this approach is computationally feasible. However, using this method only a \textit{subset} of the whole point clouds are selected, which is not desired since the user intends to receive \textit{all} the necessary point clouds. Our proposed algorithms select all necessary point clouds, however, with different bitrates according to the models' priorities given the user's view. This is a \textit{multiple-choice knapsack problem (MCKP)} in which the items (point cloud models in our context) are organized into groups corresponding to the items. Each group contains higher bitrate point clouds corresponding to a model and lower-bitrate versions of the same point cloud model given the adaptation manifest that we have designed.

\begin{figure}[!t]
\centering
\includegraphics[width=1\columnwidth]{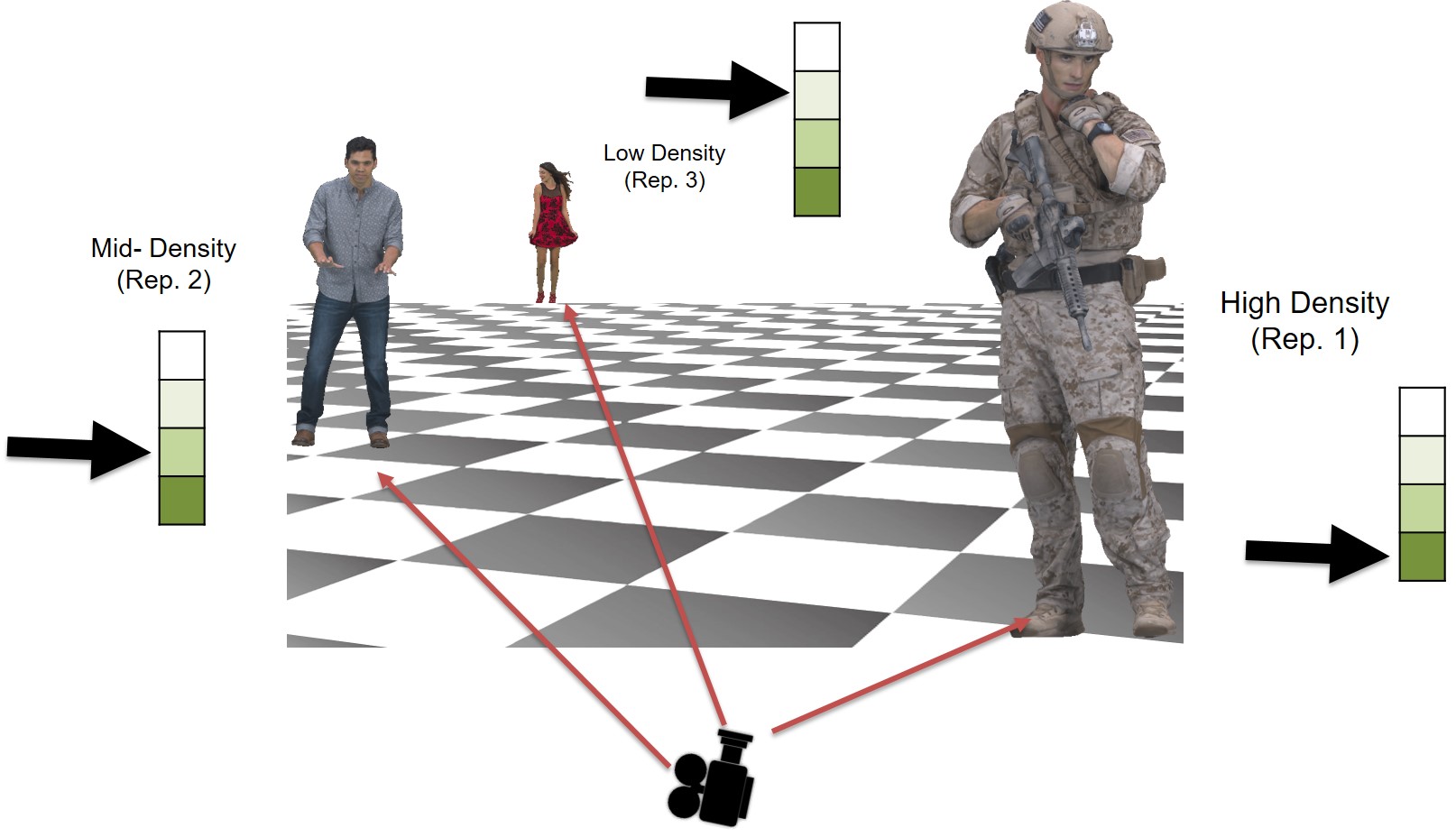}
\caption{An example point cloud prioritization in a viewport. Point clouds closer to the camera are assigned highest priority ($C_1$) and therefore higher quality representation, while point clouds farther away are assigned lowest priority ($C_3$), and therefore, lower quality representation.}
\label{fig:priority}
\end{figure}
There are $n$ point cloud models $\mathcal{T}=\{\tau_1,\tau_2,\ldots,\tau_n\}$ in the 360-degree environment. The highest possible representation of each $\tau_h \in \mathcal{T}$ has a bitrate requirement of $s_{\tau_h}$ and a priority or importance coefficient of $p_{\tau_h}$ given it's view or distance. With view awareness feature, our algorithm assigns highest priority ($C_1$) to the point clouds more important for the user's view (within the user's immediate viewport or closer), and lowest priority ($C_3$) to the point clouds (either outside of user's immediate viewport or farther away). Figure \ref{fig:priority} illustrates how our prioritization approach is applied against point cloud models inside a 3D scene. We assume the quality contribution of a point-cloud $\tau_h$ is a simple function $q_{\tau_h} = p_{\tau_h}\times s_{\tau_h}$. The available bandwidth in every interval limits the total bitrate of all point clouds that can be received at the receiving device to $W$, which serves as an available bandwidth budget. Our rate allocation heuristic is general, and can be employed based on any number of priority classes, or any type of available resources, such as energy budget or available CPU processing power. In this pilot study, we use three priority classes, and assume $W$ to be the available bandwidth.

Let $X= \{x_1,x_2,\ldots,x_n\}$, be the set of point clouds which are received at the device, serving as the output of running rate allocation algorithm. Each $x_i \in X$ corresponds to an original point cloud $\tau_i \in \mathcal{T}$. Similarly, each $x_i$ has a priority coefficient $p_{x_i} = p_{\tau_i}$ depending on the view.

We assume there are $L$ number of point clouds representations available given the manifest, with a representation of level $k$ noted as $R_k$ ($0 \leq k \leq L$) and the bitrate of a point cloud $\tau_i$ with representation $R_k$ noted as $s_{\tau_i}^{R_k}$. We assume the lowest bitrate corresponds to the representation with the highest ID (i.e. $R_L$) which is determined as the \textit{minimum bitrate} that is acceptable for a user. In a similar way, the quality contribution of a point cloud $x_i$ is $q_{x_i}=p_{x_i}\times s_{x_i}$.

\subsection{Heuristic Algorithm}

\begin{algorithm}[!t]
\begin{algorithmic}
\State $\mathcal{T}$: prioritized list of point clouds sorted from highest to lowest priority
\State $\tau_i$: point cloud with highest bitrate $s_{\tau_i}$
\State $x_i$: adapted point cloud with bitrate $s_{x_i}$
\State $L$: Number of representation levels
\State $R_L$: Level $L$ representation
\State Calculate $W_{min} = \sum s_{\tau_i}^{R_L}$ \%comment: minimum bitrate requirement for all point clouds
\State $\forall \tau_i \in \mathcal{T}: s_{x_i} \gets s_{\tau_i}^{R_L}$ \%comment: assign $R_L$ (minimum bitrate) to all $\tau_i$'s.
\State $W_0 \gets W - W_{min}$ \%comment: initialization
\While {$s_{\tau_i}- s_{\tau_i}^{R_L} \leq W_{i-1}$}
\%comment: i=1 initially.
\State $s_{x_i} \gets s_{\tau_i}$
\State $W_i \gets W_{i-1} - (s_{\tau_i}- s_{\tau_i}^{R_L})$
\State{ $i \gets i+1$~~\%comment: adapt next point cloud}
\EndWhile \\
\%comment: above loop repeats until a point cloud $\tau_{\ell}$ cannot be delivered at highest bitrate within the remaining bandwidth budget $W_{\ell-1}$.
\State $\ell \gets i$ \%comment: resulting from above loop.
\State{Find lowest $L' \leq L$ such that \\$s_{\tau_\ell}^{R_{L'}} \leq W_{\ell-1}+ s_{\tau_\ell}^{R_L}$ ~~\%comment: determines the highest bitrate possible at which $\tau_{\ell}$ can be received within remaining budget, by calculating the lowest representation level $L'$.}
\State{ $s_{x_{\ell}} \gets s_{\tau_\ell}^{R_{L'}}$ \%comment: adapt $\tau_{\ell}$ and calculate $s_{x_{\ell}}$}
\end{algorithmic}
 \caption*{Rate allocation heuristic algorithm for point cloud streaming}
\end{algorithm}

Let $S$ be the total bitrate requirement of all point clouds, and $W$ be the available bandwidth budget. The minimum quality acceptable for users is given as the representation of level $L$ noted as $R_{L}$. Let $C_{1}$, $C_{2}$, and $C_{3}$ be the class of point clouds with the highest priority, medium priority, and lowest priority, respectively.

For each point cloud $\tau_i$ in $\mathcal{T}$, we calculate $q_i$ as described previously. This is the contribution that $\tau_i$ would make to the average quality of the 3D world system if it were received at highest bitrate possible. We then calculate $W_{\min} = \sum s_{\tau_i}^{R_L}$ which is the minimum bitrate that is needed to receive all point clouds at their lowest bitrates. In the following, assume that $W_{\min} \leq W$ so the unused bitrate budget would be $W_0=W-W_{\min}$.

To determine the best bitrate for each point cloud, our algorithm sorts the prioritized list of point clouds by the global priority from the largest to the smallest. For ease of notation in the following, suppose that the point clouds are re-indexed so that the sorted list of point clouds is $\tau_1,\tau_2,\ldots,\tau_n$. If $s_{\tau_1}- s_{\tau_1}^{R_L}\leq W_0$ then there is enough unused budget to receive $\tau_1$ at highest bitrate ($R_0$), so the point cloud $x_1$ would have $s_{x_1}=s_{\tau_1}$ and would contribute $q_1$ to the average quality. This leaves an unused bandwidth budget of $W_1=W_0 - s_{\tau_1}^{R_0}- s_{\tau_1}^{R_L}$ for the remaining point clouds after $x_1$. The algorithm repeats for $\tau_2, \tau_3,\ldots$ until some point clouds $\tau_{\ell}$ cannot be received at highest bitrate within the remaining budget $W_{\ell-1}$. It then determines the highest possible bitrate at which it can be received by calculating the lowest representation level $L': L'\leq L$ such that $s_{\tau_\ell}^{R_{L'}} \leq W_{\ell-1}+ s_{\tau_\ell}^{R_{L}}$. The point cloud $x_{\ell}$ will have bitrate $s_{x_{\ell}} = s_{\tau_\ell}^{R_{L'}}$ and will contribute $q'_{\ell}$ to the average quality of the whole. The remaining bandwidth budget after streaming $x_{\ell}$ will be $W_{\ell} = W_{\ell-1} - s_{\tau_\ell}^{R_{L'}}$. The algorithm repeats this process to determine the proper bitrates, amount of bandwidth budget, and quality contribution for each of the remaining point clouds $x_{\ell+1},x_{\ell+2},\ldots,x_{n}$.

The algorithm needs a one-time implementation in the beginning of the session for the main process. Therefore it is implemented in real-time and does not provide any additional overhead during the runtime. It is implemented efficiently in $O(n log n)$ time and $O(n)$ space and produces solutions close optimal. The approximation error depends on the difference between the bitrate chosen for the first point cloud that cannot be received at highest bitrate (i.e. $\tau_\ell$) and the remaining budget available to receive it.

\section{References}
\begin{itemize}
\item M. Hosseini and C. Timmerer, "Dynamic Adaptive Point Cloud Streaming", In Proceedings of the 23rd Packet Video Workshop, 2018. ACM, New York, 1--7.
\item M. Hosseini, V. Swaminathan, "Adaptive 360 VR video streaming: Divide and conquer!", IEEE International Symposium on Multimedia (ISM), San Jose, 2016.
\item M Hosseini, Y Jiang, RR Berlin, L Sha, H Song, "Toward physiology-aware DASH: Bandwidth-compliant prioritized clinical multimedia communication in ambulances", IEEE Transactions on Multimedia, 2017.
\end{itemize}
\end{document}